\documentclass[letterpaper]{jpconf}
\usepackage{graphicx}
\newcommand{\ba}{\begin{eqnarray}}
\newcommand{\ea}{\end{eqnarray}}
\begin{document}
\title{Geometric symmetries in light nuclei}

\author{Roelof Bijker}

\address{Instituto de Ciencias Nucleares, 
Universidad Nacional Aut\'onoma de M\'exico, 
A.P. 70-543, 04510 M\'exico, D.F., M\'exico}

\ead{bijker@nucleares.unam.mx}

\begin{abstract}
The algebraic cluster model is is applied to study cluster states in the nuclei 
$^{12}$C and $^{16}$O. The observed level sequences can be understood in terms of  
the underlying discrete symmetry that characterizes the geometrical configuration 
of the $\alpha$-particles, {\it i.e.} an equilateral triangle for $^{12}$C, and a 
regular tetrahedron for $^{16}$O. The structure of rotational bands provides a 
fingerprint of the underlying geometrical configuration of $\alpha$-particles. 
\end{abstract}

\section{Introduction}

Ever since the early days of nuclear physics the structure of $^{12}$C has been 
extensively investigated both experimentally and theoretically \cite{wheeler,Teller,Oertzen,FreerFynbo}. 
In recent years, the measurement of new rotational excitations of both the ground 
state \cite{Fre07,Kirsebom,Marin} and the Hoyle state \cite{Itoh,Freer2,Gai,Freer4} 
has generated a lot of renewed interest to understand the structure of $^{12}$C and 
that of $\alpha$ cluster nuclei in general. Especially the (collective) nature of the 
$0^+$ Hoyle state at 7.65 MeV which is of crucial importance in stellar nucleosynthesis 
to explain the observed abundance of $^{12}$C, has presented a challenge to 
nuclear structure calculations, such as $\alpha$-cluster models \cite{Robson}, 
antisymmetrized molecular dynamics \cite{AMD}, fermionic molecular dynamics \cite{FMD}, 
BEC-like cluster model \cite{BEC}, (no-core) shell models \cite{Roth,Draayer}, 
{\it ab initio} calculations based on lattice effective field theory \cite{Epelbaum1,Epelbaum2}, 
and the algebraic cluster model \cite{Marin,ACM,O16}. 

In this contribution, I discuss some properties of the $\alpha$-cluster nuclei $^{12}$C 
and $^{16}$O in the framework of the algebraic cluster model. 

\section{Algebraic Cluster Model}

The Algebraic Cluster Model (ACM) describes the relative motion of the $n$-body clusters 
in terms of a spectrum generating algebra of $U(\nu+1)$ where $\nu=3(n-1)$ represents the 
number of relative spatial degrees of freedom. 
For the two-body problem the ACM reduces to the $U(4)$ vibron model \cite{cpl}, for 
three-body clusters to the $U(7)$ model \cite{ACM,BIL} and for four-body clusters 
to the $U(10)$ model \cite{O16,RB}. 
In the application to $\alpha$-cluster nuclei the Hamiltonian has to be 
invariant under the permuation group $S_n$ for the $n$ identical $\alpha$ particles. 
Since one does not consider the excitations of the $\alpha$ particles themselves,  
the allowed cluster states have to be symmetric under the permutation group. 

\begin{table}
\centering
\caption{Algebraic Cluster Model for two-, three- and four-body clusters} 
\label{ACMsummary}
\vspace{5pt}
\begin{tabular}{cccc}
\hline
\hline
\noalign{\smallskip}
& $2\alpha$ & $3\alpha$ & $4\alpha$ \\
\noalign{\smallskip}
\hline
\noalign{\smallskip}
ACM & $U(4)$ & $U(7)$ & $U(10)$ \\
Point group & ${\cal C}_2$  & ${\cal D}_{3h}$ & ${\cal T}_{d}$ \\
Geometry & Linear & Triangle & Tetrahedron \\
\noalign{\smallskip}
\hline
\noalign{\smallskip}
G.s. band & $0^+$ & $0^+$ & $0^+$ \\
          & $2^+$ & $2^+$ &       \\
          &       & $3^-$ & $3^-$ \\
          & $4^+$ & $4^{\pm}$ & $4^+$ \\
          &       & $5^-$ &       \\
          & $6^+$ & $6^{\pm+}$ & $6^{\pm}$ \\  
\noalign{\smallskip}
\hline
\hline
\end{tabular}
\end{table}

The potential energy surface corresponding to the $S_n$ invariant ACM Hamiltonian gives 
rise to several possible equilibrium shapes. In addition to the harmonic oscillator (or 
$U(3n-3)$ limit) and the deformed oscillator (or $SO(3n-2)$ limit), there are other 
solutions which are of special interest for the applications to $\alpha$-cluster nuclei. 
These cases correspond to a geometrical configuration of $\alpha$ particles located at the 
vertices of an equilateral triangle for $^{12}$C and of a regular tetrahedron for $^{16}$O. 
Even though they do not correspond to dynamical symmetries of the ACM Hamiltonian, one can 
still obtain approximate solutions for the rotation-vibration spectrum  
\ba
E &=& \left\{ \begin{array}{lll} 
\omega_{1}(v_{1}+\frac{1}{2}) + \omega_{2}(v_{2}+1) + \kappa \, L(L+1) && \mbox{ for $n=3$} \\ \\
\omega_{1}(v_{1}+\frac{1}{2}) + \omega_{2}(v_{2}+1) + \omega_{3}(v_{3}+\frac{3}{2}) + \kappa \, L(L+1) && \mbox{ for $n=4$} 
\end{array} \right. 
\label{energy} 
\nonumber
\ea
The rotational structure of the ground-state band depends on the point group symmetry of the 
geometrical configuration of the $\alpha$ particles and is summarized in Table~\ref{ACMsummary}. 

\begin{figure}
\begin{minipage}{.5\linewidth}
\includegraphics[width=\linewidth]{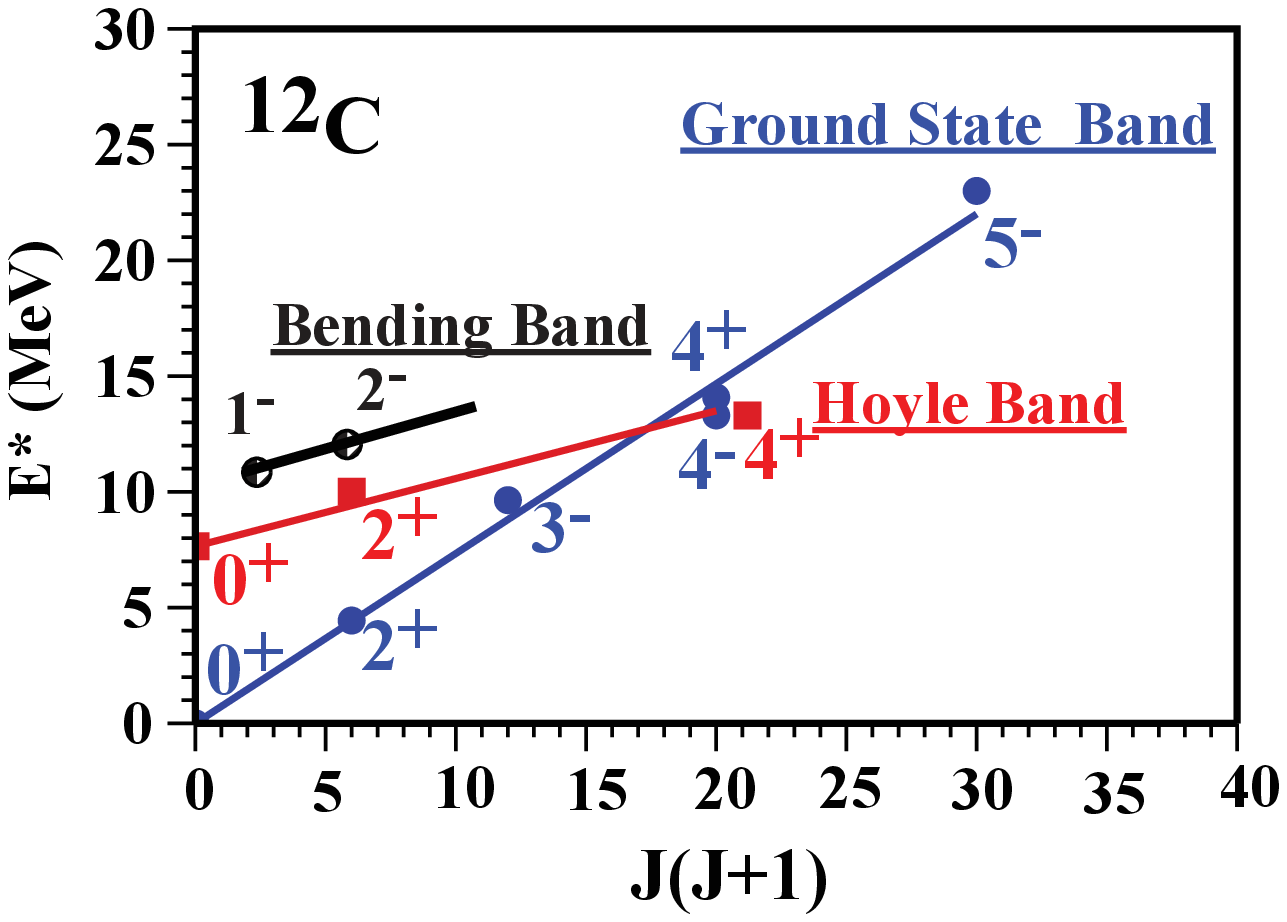}
\end{minipage}\hfill
\begin{minipage}{.5\linewidth}
\includegraphics[width=\linewidth]{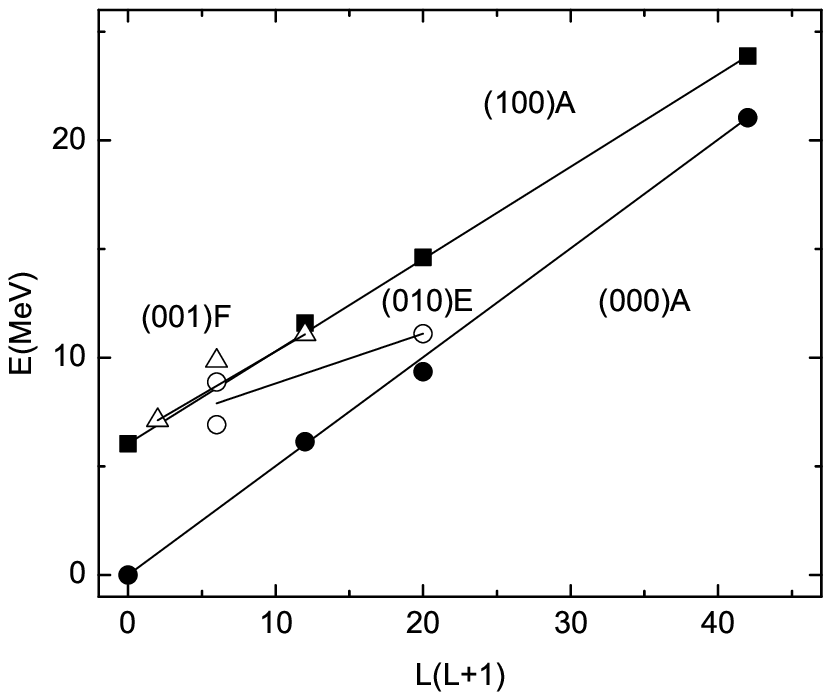}
\end{minipage}\hfill
\caption{(Color online) Rotational band structure of the ground-state band, the Hoyle band (or $A$ 
vibration) and the bending vibration (or $E$ vibration) in $^{12}$C (left) \cite{Marin}, 
and the ground-state band (closed circles), the $A$ vibration (closed squares), the $E$ vibration 
(open circles) and the $F$ vibration (open triangles) in $^{16}$O (right) \cite{O16}.}
\label{RotBands}
\end{figure}

The triangular configuration with three $\alpha$ particles has point group symmetry ${\cal D}_{3h}$ 
\cite{ACM}. Since ${\cal D}_{3h} \sim {\cal D}_{3} \times P$, the transformation properties under 
${\cal D}_{3h}$ are labeled by parity $P$ and the representations of ${\cal D}_{3}$ which is 
isomorphic to the permutation group $S_{3}$. The corresponding rotation-vibration spectrum is that 
of an oblate top: $v_{1}$ represents the vibrational quantum number for a symmetric stretching $A$ 
vibration, $v_2$ denotes a doubly degenerate $E$ vibration. The rotational band structure of $^{12}$C 
is shown in the left panel of Fig.~\ref{RotBands}. 

The tetrahedral group ${\cal T}_d$ is isomorphic to the permutation group $S_4$. In this case, there 
are three fundamental vibrations: $v_{1}$ represents the vibrational quantum number for a symmetric 
stretching $A$ vibration, $v_2$ denotes a doubly degenerate $E$ vibration, and $v_3$ a three-fold 
degenerate $F$ vibration. The right panel of Fig.~\ref{RotBands} shows the rotational band structure 
of $^{16}$O. 

\section{Electromagnetic transitions}

\begin{table}
\centering
\caption{$B(EL)$ values in $^{12}$C (top) and $^{16}$O (bottom).}
\label{bem}
\vspace{5pt}
\begin{tabular}{cccll}
\hline
\noalign{\smallskip}
$^{12}$C & Th & Exp & & Ref \\
\noalign{\smallskip}
\hline
\noalign{\smallskip}
$B(E2;2_{1}^{+} \rightarrow 0_{1}^{+})$ &  8.4 & $7.6 \pm 0.4$ & $e^{2}\mbox{fm}^{4}$ & \cite{ajz,reuter,strehl} \\
$B(E3;3_{1}^{-} \rightarrow 0_{1}^{+})$ & 44   & $103 \pm 17$ & $e^{2}\mbox{fm}^{6}$ & \cite{ajz,reuter,strehl} \\
$B(E4;4_{1}^{+} \rightarrow 0_{1}^{+})$ & 73   &              & $e^{2}\mbox{fm}^{8}$ & \cite{ajz,reuter,strehl} \\  
$M(E0;0_{2}^{+} \rightarrow 0_{1}^{+})$ &  0.4 & $5.5 \pm 0.2$ & $\mbox{fm}^{2}$ & \cite{ajz,reuter,strehl} \\
\noalign{\smallskip}
\hline
\noalign{\smallskip}
$^{16}$O & Th & Exp & & Ref \\
\noalign{\smallskip}
\hline
\noalign{\smallskip}
$B(E3;3_1^- \rightarrow 0_1^+)$ &  215 & $205 \pm  10$ & $e^{2}\mbox{fm}^{6}$ & \cite{NDS} \\
$B(E4;4_1^+ \rightarrow 0_1^+)$ &  425 & $378 \pm 133$ & $e^{2}\mbox{fm}^{8}$ & \cite{NDS} \\
$B(E6;6_1^+ \rightarrow 0_1^+)$ & 9626 &               & $e^{2}\mbox{fm}^{12}$ & \cite{NDS} \\ 
$M(E0;0_{2}^{+} \rightarrow 0_{1}^{+})$ &  0.54 & $3.55 \pm 0.21$ & $\mbox{fm}^{2}$ & \cite{NDS} \\
\noalign{\smallskip}
\hline
\end{tabular}
\end{table}

For transitions along the ground state band the transition form factors are given in terms of a 
product of a spherical Bessel function and an exponential factor arising from a Gaussian distribution 
of the electric charges, ${\cal F}(0^+ \rightarrow L^P;q) = c_L \, j_L(q \beta) \, \mbox{e}^{-q^{2}/4\alpha}$ 
\cite{ACM}. The charge radius can be obtained from the slope of the elastic form 
factor in the origin $\langle r^{2} \rangle^{1/2} = \sqrt{\beta^{2}+3/2\alpha}$. 
The transition form factors depend on the parameters $\alpha$ and $\beta$ which can be determined from 
the first minimum in the elastic form factor and the charge radius. 

The transition probabilities $B(EL)$ along the ground state band can be extracted 
from the form factors in the long wavelength limit 
\ba
B(EL;0^+ \rightarrow L^P) \;=\; \frac{(Ze)^2}{4\pi} \, c_L^2 \, \beta^{2L} ~, 
\nonumber
\ea
with
\ba
c_L^2 = \left\{ \begin{array}{lcl} 
\frac{2L+1}{3} \left[ 1+2P_{L}(-\frac{1}{2}) \right] && \mbox{for $n=3$} \\ \\
\frac{2L+1}{4} \left[ 1+3P_{L}(-\frac{1}{3}) \right] && \mbox{for $n=4$} \end{array} \right.
\nonumber
\ea
The good agreement for the $B(EL)$ values for the ground band in Table~\ref{bem} shows that both in 
$^{12}$C and in $^{16}$O the positive and negative parity states merge into a single rotational band.   
Moreover, the large values of $B(EL;L_{1}^{P} \rightarrow 0_{1}^{+})$ indicate a collectivity which is not 
predicted for simple shell model states. The large deviation for the $E0$ between the first excited $0^+$ 
(Hoyle) state and the ground state indicates that the $0^{+}_2$ state cannot be interpreted as a 
simple vibrational excitation of a rigid triangular ($^{12}$C) or tetrahedral ($^{16}$O) configuration, 
but rather corresponds to a more floppy configuration with large rotation-vibration couplings. 
A more detailed study of the electromagnetic properties of $\alpha$-cluster nuclei in the ACM for non-rigid 
configurations is in progress. 

\section{Summary and conclusions}

In this contribution, the cluster states in $^{12}$C and $^{16}$O were interpreted 
in the framework of the ACM as arising from the rotations and vibrations of a triangular 
and tetrahedral configuration of $\alpha$ particles, respectively. 
In both cases, the ground state band consist of positive and negative parity states 
which coalesce to form a single rotational band. This interpretation is validated by 
the observance of strong $B(EL)$ values. The rotational sequences can be considered 
as the fingerprints of the underlying geometric configuration (or point-group symmetry) 
of $\alpha$ particles.  

For the Hoyle band in $^{12}$C there are several interpretations for the geometrical 
configuration of the three $\alpha$ particles. In order to determine whether the geometrical 
configuration of the $\alpha$-particles for the Hoyle band is linear, bent or triangular, the 
measurement of a possible $3^-$ Hoyle state is crucial, since its presence would indicate  
a triangular configuration, just as for the ground state band.  

Finally, the results presented here for $^{12}$C and $^{16}$O emphasize the occurrence 
of $\alpha$-cluster states in light nuclei with ${\cal D}_{3h}$ and ${\cal T}_d$ point 
group symmetries, respectively.  

\ack
This work was supported in part by research grant IN107314 from PAPIIT-DGAPA.

\end{document}